\documentclass[preprint,aps,prd,nofootinbib]{revtex4}
\usepackage[dvips]{graphicx}
\begin{document}

\preprint{CALT-68-2450}
\preprint{hep-ph/0310025}
\title{Perturbative corrections to curvature sum rules}
\author{Matthew P.\ Dorsten}
\email{dorsten@theory.caltech.edu}
\affiliation{California Institute of Technology, Pasadena, CA 91125
        \\ $\phantom{}$ }

\vspace*{1.5cm}

\begin{abstract}
Two new sum rules were recently discovered by Le Yaouanc \emph{et
al.}\ by applying the operator product expansion to the nonforward
matrix element of a time-ordered product of $b \to c$ currents in the
heavy-quark limit of QCD\@. They lead to the constraints
$\sigma^2>5\rho^2/4$ and $\sigma^2>3(\rho^2)^2/5+4\rho^2/5$ on the
curvature of the $\bar{B} \to D^{(*)}$ Isgur-Wise function, both of
which imply the absolute lower bound $\sigma^2>15/16$ when combined
with the Uraltsev bound $\rho^2>3/4$ on the slope. This paper
calculates order $\alpha_s$ corrections to these bounds, increasing
the accuracy of the resultant constraints on the physical form
factors. The latter may have implications for the determination of
$|V_{cb}|$ from exclusive semileptonic $B$ meson decays.
\end{abstract}

\maketitle

\newpage

\section{Introduction}

Heavy quark effective theory (HQET)~\cite{hqet} provides a
model-independent method of extracting the CKM matrix element
$|V_{cb}|$ from exclusive semileptonic $B$ meson decays. The
$\bar{B}\to D^{(*)}l\bar{\nu}$ differential decay rates are given by
\begin{eqnarray}
\frac{\mathrm{d}\Gamma}{\mathrm{d}w}
(\bar{B} \to D^* l \bar{\nu})&=&
\frac{G^2_F |V_{cb}|^2 m_B^5}{48 \pi^3}\,r_*^3 (1-r_*)^2
\sqrt{w^2-1}\,(w+1)^2 \nonumber \\
&& \times \left[1+\frac{4w}{w+1}\frac{1-2wr_*+r_*^2}{(1-r_*)^2}\right]
\mathcal{F}_{D^*}(w)^2 \nonumber \ , \\
\frac{\mathrm{d}\Gamma}{\mathrm{d}w}
(\bar{B} \to D l \bar{\nu})&=&
\frac{G^2_F |V_{cb}|^2 m_B^5}{48 \pi^3}\,r^3 (1+r)^2
(w^2-1)^{3/2}\mathcal{F}_D(w)^2 \ ,
\end{eqnarray}
where $r_{(*)}=m_{D^{(*)}}/m_B$ and $w=v \cdot v'$ is the product of
the $\bar{B}$ and $D^{(*)}$ four-velocities. Heavy quark
symmetry~\cite{iw} relates $\bar{B} \to D^{(*)} l \bar{\nu}$ form
factors to the corresponding Isgur-Wise function, with the result
$\mathcal{F}_{D^*}(w)=\mathcal{F}_D(w)=\xi(w)$ in the heavy-quark
limit of QCD\@. Since $\xi(w)$ is absolutely normalized to unity at
zero recoil (i.e., $w=1$)~\cite{iw,nus,shif,luke}, experimental data
determine $|V_{cb}|$ without recourse to model-specific assumptions.

This procedure has several sources of uncertainty. First, the identity
$\mathcal{F}_{D^{(*)}}(1)=1$ receives both perturbative corrections
and corrections suppressed by the heavy $b$ and $c$ quark masses. The
former are known to order $\alpha_s^2$~\cite{czar}, with unknown
higher-order corrections likely less than 1\%, but the latter depend
on four subleading Isgur-Wise functions and have been estimated only
with phenomenological models and quenched lattice QCD\@.

Another source of error is the extrapolation of measured form factors
to zero recoil, where the rates vanish. Linear fits of
$\mathcal{F}_{D^{(*)}}(w)$ underestimate the zero-recoil value by
about 3\%, an effect mostly due to the curvature~\cite{stone}. Using
non-linear shapes for $\mathcal{F}_{D^{(*)}}(w)$ reduces this error,
and therefore constraints on second and higher derivatives at
zero-recoil are welcome. Dispersive constraints~\cite{disper1,disper2}
relate second and sometimes higher derivatives to the first and are
commonly used.

HQET sum rules provide a complementary way of constraining the
$\mathcal{F}_{D^{(*)}}(w)$ shapes. New sum rules for the curvature and
higher derivatives of the Isgur-Wise function were derived in
Refs.~\cite{ley1,ley2,ley3}. Equating the result of inserting a
complete set of intermediate states in the nonforward matrix element
of a time-ordered product of HQET currents with the operator product
expansion (OPE) gives a generic sum rule depending on the products of
the velocities of the initial, final, and intermediate states. These
are denoted respectively by $v_i$, $v_f$, and $v'$ (the intermediate
states all have the same velocity $v'$ in the infinite-mass limit),
and the products are denoted by
\begin{equation}
\begin{array}{r @{,} c @{,} l}
w_{if}=v_i \cdot v_f \quad
& \quad w_i=v_i \cdot v' \quad 
& \quad w_f=v_f \cdot v' \, ,
\end{array}
\end{equation}
or generically $w_x$. These parameters are constrained to lie within
the range~\cite{ley1}
\begin{equation}
w_i,w_f,w_{if} \geq 1 \ , \quad
w_i w_f-\sqrt{(w_i^2-1)(w_f^2-1)} \leq w_{if} \leq
w_i w_f+\sqrt{(w_i^2-1)(w_f^2-1)} \ ,
\end{equation}
and differentiating the generic sum rule with respect to them at
$w_x=1$ (read: $w_i=w_f=w_{if}=1$) produces a class of sum rules for
derivatives of the Isgur-Wise function at zero recoil. The sum rules
of Refs.~\cite{ley1,ley2,ley3} were derived at tree level in the
heavy-quark limit. The present paper includes the order $\alpha_s$
corrections to the new sum rules and uses them to derive bounds on the
curvatures of $\mathcal{F}_{D^{(*)}}(w)$ including $\alpha_s$ and
$\Lambda_{\mathrm{QCD}}/m_{c,b}$ corrections. Including these
corrections to the heavy-quark limit is important for meaningful
comparison with data and dispersive constraints.

\section{Derivation of the Generic Sum Rule}
\label{tw}

The derivation of the generic sum rule follows a well-known
formalism~\cite{bigi,kap,boyd,bjo}. It begins with the consideration
of the time-ordered product of two arbitrary heavy-heavy currents
\begin{eqnarray}
\label{tfi}
T_{fi}(\varepsilon)&=&\frac{i}{2 m_B} \int d^4x \ e^{-iq \cdot x}\, 
\langle B_f(p_f)|\,T\{J_{\!f}(0),J_{i}(x)\}\,|B_i(p_i) \rangle
\end{eqnarray}
as a complex function of $\varepsilon=E_M-E_i-q^0$ at fixed $\vec{q}$,
where $E_M=\sqrt{m^2_M+|\vec{p}_i+\vec{q}\,|^2}$ is the minimum
possible energy of the charmed hadronic state that $J_i$ can create at
fixed $\vec{q}$. The currents have the form 
\begin{equation} 
J_{\!f}(x) = \bar{b}(x) \Gamma_{\!\!f} c(x) \ ,\
J_i(y) = \bar{c}(y) \Gamma_{\!i} b(y)
\end{equation}
for any Dirac matrices $\Gamma_{\! i,f}$. Only the choices $\Gamma_{\!
i,f}={/ \hskip - 0.53em v}_{i,f}$ and $\Gamma_{\! i,f}={/
\hskip - 0.53em v}_{i,f} \gamma_5$ are explored here, but in principle
others lead to different sum rules. The $B$ states are ground state
$\bar{B}$ or $\bar{B}^*$ mesons and have the standard relativistic
normalization. As in the derivation of the Uraltsev sum
rule~\cite{uraltsev}, the initial and final states do not necessarily
have the same velocity. Considering the \emph{nonforward} matrix
element of the time-ordered product is a crucial generalization in
deriving the new sum rules~\cite{ley1}.

From Eq.~(\ref{tfi}) one proceeds by splitting up the two
time-orderings and inserting complete sets of intermediate charm
states. The result is
\begin{eqnarray}
T_{fi}(\varepsilon) & = & \frac{1}{2 m_B} \sum_{X_c} (2 \pi)^3 
\delta^3(\vec{q}+\vec{p}_i-\vec{p}_{X_c}) 
\frac{\langle B_f|J_{\!f}(0)|X_c\rangle\langle X_c|J_i(0)|B_i\rangle}
{\varepsilon+E_{X_c}-E_M-i0^+} \nonumber \\
&& \mbox{} -\frac{1}{2 m_B} \sum_{X_{\bar{c}bb}}(2\pi)^3 
\delta^3(\vec{q}-\vec{p}_f+\vec{p}_{X_{\bar{c}bb}}) 
\frac{\langle B_f|J_i(0)|X_{\bar{c}bb}\rangle\langle X_{\bar{c}bb}|
J_{\!f}(0)|B_i\rangle}{\varepsilon+E_i+E_f-E_M-E_{X_{\bar{c}bb}}+i0^+}\ ,
\end{eqnarray}
where the sums include phase space factors such as $d^3p/(2\pi)^3
2E_X$. Again, $T_{fi}$ has been written as above to call attention to
the full generality possible for deriving sum rules by this method,
but here both $B_i$ and $B_f$ will be taken to be $\bar{B}$ mesons to
avoid the considerable complication of the $\bar{B}^*$
polarization. In addition, HQET states and currents will be used
henceforth since the goal is sum rules for the derivatives of the
Isgur-Wise function. Deriving the bounds in the effective theory also
makes the calculation of perturbative corrections much easier.

\begin{figure}[!ht]
\begin{center}
\includegraphics[width=0.6\textwidth]{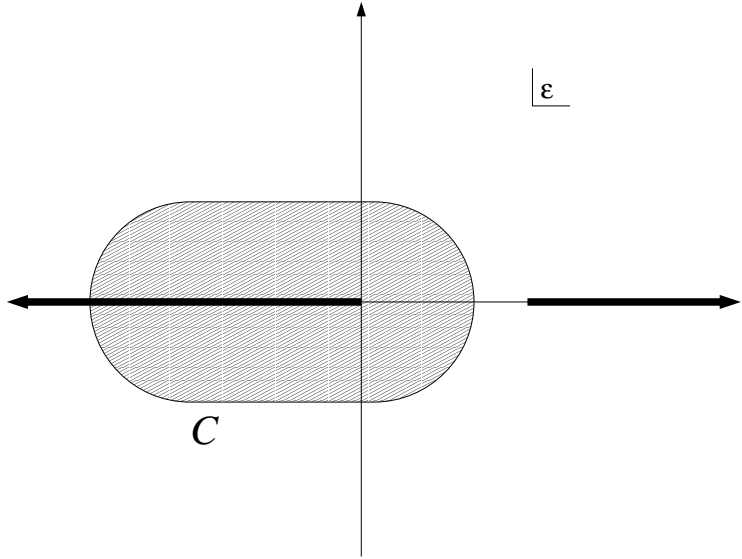}
\caption{\small The cuts of $T_{fi}$ in the complex $\varepsilon$
plane. The depicted contour picks up only contributions from the
left-hand cut, which corresponds to physical states with a charm
quark. The states given by the right-hand cut do not contribute here.}
\label{contour}
\end{center}
\end{figure}

The function $T_{fi}(\varepsilon)$ has two cuts along the real axis,
as shown in Fig.~\ref{contour}. The important one here, running from
$-\infty$ to the origin, comes from the first time-ordering and
corresponds to intermediate states with a $c$ quark or a $c$ quark, a
$b$ quark, and a $\bar{b}$ quark. The cut associated with the other
time-ordering begins near $2 m_c$ and corresponds to states with two
$b$ quarks and a $\bar{c}$ quark. Since $T_{fi}(\varepsilon)$ is
perturbatively calculable only when smeared over a large enough range
of $\varepsilon$~\cite{wein}, it is multiplied by a weight function
$W_{\!\Delta}$ and integrated around the contour shown in the
figure. The scale $\Delta$ gives the extent of the smearing and
therefore should be well above $\Lambda_{\mathrm{QCD}}$. The contour
chosen eliminates all but the intermediate states with heavy quark
content $c$ by avoiding the second cut and pinching the first at
$\varepsilon=-2 m_b$. Crossing the contour assumes local duality at
the scale $m_b$, but if $\Delta < m_b$ the weight function will be
quite small here. This will be clear with the specific weight function
used below.
Assuming it is analytic in the shaded region of Fig.~\ref{contour},
the result is
\begin{eqnarray}
\label{preope}
\frac{1}{2\pi i}\int_C d\varepsilon \, W_{\!\Delta}(\varepsilon)\,
T_{fi}(\varepsilon) & = & 
\sum_{X_c} W_{\!\Delta}(E_M\! -\! E_{X_c})
\frac{\langle \bar{B}(v_f)|J_{\!f}|X_c(v') \rangle
\langle X_c(v')|J_i|\bar{B}(v_i)\rangle}
{4 v'^0}\ ,
\end{eqnarray}
where the delta function has been used to perform the phase-space
integral and the HQET state normalization convention has been used to
eliminate mass factors in the denominator. The intermediate $X_c$
states carry four-momentum $p_{X_c}=m_{X_c}v'=p_i+q$.

Choice of the weight function is governed by well-known
concerns~\cite{bjo,kap}. In practice one uses
$W_{\!\Delta}(\varepsilon)=\theta(\Delta+\varepsilon)$, which is the $n
\to \infty$ limit of the set of functions
\begin{equation}
W^{(n)}_{\!\Delta}(\varepsilon)=\frac{\Delta^{2n}}{\varepsilon^{2n}+\Delta^{2n}}
\end{equation}
for $\varepsilon < 0$. But since the weight function must be analytic
within the contour, the use of these is strictly correct only for
small $n>1$. In this case the poles at $\varepsilon =
\sqrt[2n]{-1}\,\Delta$ are a distance of order $\Delta$ away from the
cut, and the contour can be deformed away without getting too close to
the cut and relying on local duality at a scale below $m_b$. This is
not true of the $n \to \infty$ limit, in which the poles approach the
cut and the contour must pinch the cut at the scale $\Delta$ instead
of $m_b$. This is a problem because the contribution at $\Delta$ is
weighted much more heavily than that at $m_b$, and thus the results
will depend more strongly on the assumption of local duality. However,
an explicit calculation shows that the results here do not depend on
$n$, just as the authors of Ref.~\cite{bjo} found in their derivation
of corrections to the Bjorken sum rule. This is not true in other
cases, such as the Voloshin sum rule~\cite{bjo}. The weight function
in what follows can therefore be considered a simple step function
excluding states with excitation energies greater than
$\Delta$. Although increasing $\Delta$ includes more states and
weakens the bounds, the cutoff energy must be chosen large enough to
make perturbative QCD appropriate. Choosing $\Delta \gtrsim 2$ GeV
should therefore be sufficient.

\begin{figure}[t]
\begin{center}
\includegraphics[bb=55 635 580 725, clip, width=\textwidth]{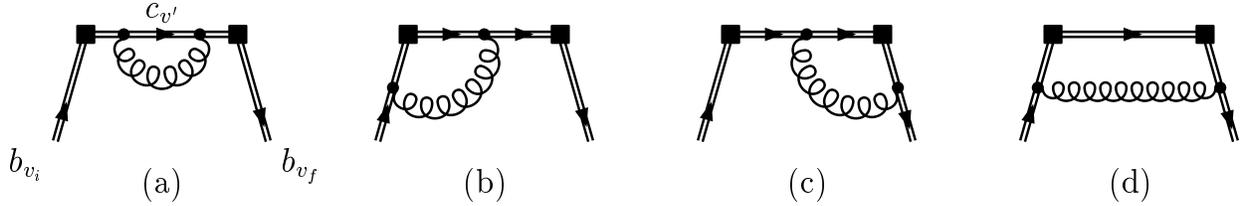}
\caption{\small Diagrams contributing to the order $\alpha_s$
corrections to the sum rules. The squares indicate insertions of the
currents $J_i$ and $J_{\!f}$, respectively. The current $J_i$ inserts
momentum $q$, while the current $J_{\!f}$ carries away momentum $q'$
sufficient to leave the final $b$-quark with velocity $v_f$. The
velocity-labeled quark fields are those of the heavy quark effective
theory.}
\label{loop}
\end{center}
\end{figure}

\begin{figure}[t]
\begin{center}
\includegraphics[bb=55 645 580 725, clip, width=\textwidth]{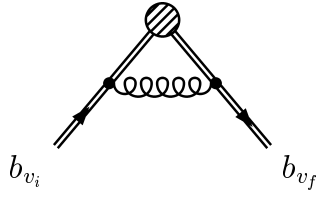}
\caption{\small One-loop renormalization of the leading operator in
the operator product expansion of $T_{fi}$. The blob indicates an
insertion of this operator, $\bar{b}_{v_f}\Gamma_{\!\!f} (1+{/ \hskip
- 0.53em v}\,')\Gamma_{\!i} b_{v_i}$. The external lines are bottom
quarks in the heavy quark effective theory.}
\label{match}
\end{center}
\end{figure}

The sum rule is derived by performing an operator product expansion on
the time-ordered product of currents on the left-hand side of
Eq.~(\ref{preope}) while writing out the right-hand side explicitly in
terms of excited-state Isgur-Wise functions. The leading-order OPE
relevant for $B$ matrix elements consists of a single dimension-three
operator, $\bar{b}_{v_f}\Gamma_{\!\!f} (1+{/ \hskip - 0.53em
v}\,')\Gamma_{\!i} b_{v_i}$. Higher dimension operators will be
neglected here, as they give corrections suppressed by powers of
$\Lambda_{\mathrm{QCD}}/\Delta$ or
$\Lambda_{\mathrm{QCD}}/m_{c,b}$. The order $\alpha_s$ corrections to
the Wilson coefficient of the leading operator are given by a matching
calculation involving the diagrams in Figs.~\ref{loop} and
\ref{match}.
The generic sum rule resulting from this is
\begin{eqnarray}
\label{generic}
\lefteqn{\frac{1}{4} \xi(w_{if})[1+ \alpha_s F(w_i,w_f,w_{if})] \,
\mathrm{Tr}\left[(1+{/ \hskip - 0.53em v}_{\! f})\, \Gamma_{\! \! f}
(1+{/ \hskip - 0.53em v}\,') \, \Gamma_{\! i}
(1+{/ \hskip - 0.53em v}_i)\right]} \nonumber \\*
& & = \sum_{X_c} W_{\!\Delta}(E_M\! -\! E_{X_c})
\langle\bar{B}(v_f)|J_{\!f}|X_c(v')\rangle
\langle X_c(v')|J_i|\bar{B}(v_i) \rangle \ ,
\end{eqnarray}
where the function $F$ contains the one-loop corrections. In
principle, $F$ could be defined to include perturbative corrections of
all orders. The form of the corrected sum rule would be the same since
HQET vertices are spin independent. The right-hand side is written out
explicitly in the next section.

Working in the rest frame of the intermediate hadrons (i.e., $v'^0=1$)
and using the $\overline{\mathrm{MS}}$ scheme with dimensional
regularization and a finite gluon mass $m$, the contributions to
$\alpha_s F$ of the graphs in Figs.~\ref{loop}(a)--\ref{loop}(d) are,
respectively,
\begin{equation}
\frac{2\alpha_s}{3\pi}\left(2-\mathrm{ln}\frac{4\Delta^2}{\mu^2}\right)\ ,
\end{equation}
\begin{equation}
\frac{2\alpha_s}{3\pi}w_i\left\{
\frac{\mathrm{ln}\left(w_i+\sqrt{w_i^2-1}\right)}{\sqrt{w_i^2-1}}
\,\mathrm{ln}\frac{4\Delta^2}{\mu^2}+\int_0^1 dx\,
\frac{2\,\mathrm{ln}x-\mathrm{ln}[1+2x(1-x)(w_i-1)]}
{1+2x(1-x)(w_i-1)}\right\}\ ,
\end{equation}
\begin{equation}
\frac{2\alpha_s}{3\pi}w_f\left\{
\frac{\mathrm{ln}\left(w_f+\sqrt{w_f^2-1}\right)}{\sqrt{w_f^2-1}}
\,\mathrm{ln}\frac{4\Delta^2}{\mu^2}+\int_0^1 dx\,
\frac{2\,\mathrm{ln}x-\mathrm{ln}[1+2x(1-x)(w_f-1)]}
{1+2x(1-x)(w_f-1)}\right\}\ ,
\end{equation}
\begin{equation}
-\frac{4\alpha_s}{3\pi}w_{if}\int_0^1 dx\,dy\,dz\,\delta(x+y+z-1)
\frac{\theta(z-\sqrt{a}m/\Delta)}{a\sqrt{z^2-am^2/\Delta^2}}\ ,
\end{equation}
where $a=1+2xy(w_{if}-1)+2xz(w_i-1)+2yz(w_f-1)$, and $\alpha_s$ is
evaluated at subtraction point $\mu$. The contribution of
Fig.~\ref{loop}(d) cannot easily be simplified further, but this is no
limitation since the sum rules require only the first few terms of $F$
in an expansion about $w_x=1$. The graph in Fig.~\ref{match}
contributes with a minus sign to the matching calculation for the
Wilson coefficient, since it gives the renormalization of the leading
operator in the OPE, and so its contribution to $\alpha_s F$ is
\begin{equation}
\frac{2\alpha_s}{3\pi}\frac{w_{if}}{\sqrt{w_{if}^2-1}}\,\mathrm{ln}
\left(w_{if}+\sqrt{w_{if}^2-1}\right)\mathrm{ln}\frac{\mu^2}{m^2}\ .
\end{equation}
This infrared divergence cancels that of the graph in
Fig.~\ref{loop}(d), leaving $\alpha_s F$ independent of the regulating
gluon mass.

From the results above, it is not hard to show that
$F(1,w,w)=F(w,1,w)=0$. This important characteristic of the
perturbative corrections is true to all orders in $\alpha_s$, as can
easily be seen. In the limits $v_i=v'$ and $v_f=v'$, one of the
currents in the time-ordered product is a conserved current associated
with heavy quark flavor symmetry and its matrix elements receive no
perturbative corrections. Because HQET loop graphs do not change the
matrix structures of inserted operators, perturbative corrections to
matrix elements of the other current cancel those of the leading
operator in the OPE\@. Therefore, the function analogous to $F$
including perturbative corrections of all orders will still vanish in
these limits.

The sum rules derived here are primarily of interest near zero recoil,
making it convenient to expand $F$ about $w_x=1$ with the definitions
\begin{eqnarray}
\alpha_s F(w_i,w_f,w_{if})&=&b_{if1}(w_{if}-1)+b_{i1}(w_i-1)
+b_{f1}(w_f-1) + \frac{1}{2}b_{if2}(w_{if}-1)^2 \nonumber \\
&&\mbox{}+\frac{1}{2}b_{i2}(w_i-1)^2 + \frac{1}{2}b_{f2}(w_f-1)^2
+ b_{i2f}(w_i-1)(w_f-1)\nonumber \\
&&\mbox{}+b_{i2if}(w_i-1)(w_{if}-1)+b_{f2if}(w_f-1)(w_{if}-1)+\cdots \ .
\end{eqnarray}
There is no zeroth-order term since $F(1,1,1)=0$. This follows from
the identities $F(1,w,w)=F(w,1,w)=0$, which also imply
\begin{eqnarray}
\label{identb1}
&b_{f1}+b_{if1}=b_{i1}+b_{if1}=0\ ,&\\
&b_{f2}+2b_{f2if}+b_{if2}=b_{i2}+2b_{i2if}+b_{if2}=0\ .&
\end{eqnarray}
These relations between derivatives of the perturbative corrections
can be checked at order $\alpha_s$ with the explicit values
\begin{eqnarray}
\label{twoderiv}
b_{if1}=-b_{f1}=-b_{i1}&=&\frac{4\alpha_s}{9\pi}\left(\frac{5}{3}-
\mathrm{ln}\frac{4\Delta^2}{\mu^2}\right)\ ,\nonumber\\
b_{i2}=b_{f2}&=&\frac{4\alpha_s}{15\pi}\left(\frac{2}{5}-
\mathrm{ln}\frac{4\Delta^2}{\mu^2}\right)\ ,\nonumber \\
b_{if2}&=& \frac{4\alpha_s}{15\pi}\left(-\frac{12}{5}+
\mathrm{ln}\frac{4\Delta^2}{\mu^2}\right)\ ,\nonumber \\
b_{i2if}=b_{f2if}&=&\frac{4\alpha_s}{15\pi}\ ,\nonumber \\
b_{i2f}&=& -\frac{4\alpha_s}{45\pi}\ .
\end{eqnarray}

The derivatives above are all specific to the rest frame of
intermediate hadrons. This is the frame used henceforth. In other
frames (e.g., $v_i^0=1$) the weight function depends on the $w_x$
parameters, and the sum rules are more complicated but not
qualitatively different.

\section{Vector and Axial Vector Sum Rules}
\label{th}

When specific matrices $\Gamma_{\! i,f}$ are chosen, the generic sum
rule in Eq.~(\ref{generic}) can be written out explicitly in terms of
excited-state Isgur-Wise functions using Falk's description of HQET
states of arbitrary spin~\cite{falk}. The choice $\Gamma_{\! i,f}={/
\hskip - 0.53em v}_{i,f}$ yields~\cite{ley2}
\begin{eqnarray}
\label{vector}
&& \xi(w_{if}) \left[1+ \alpha_s F(w_i,w_f,w_{if})\right]
(1+w_i+w_f+w_{if}) \nonumber \\
&=& (w_i+1)(w_f+1)\!\sum_{\ell=0}\frac{\ell+1}{2\ell+1} 
S_{\ell}(w_i,w_f,w_{if})\!\sum_{n}\tau_{\ell+1/2}^{(\ell)(n)}(w_i)
\tau_{\ell+1/2}^{(\ell)(n)}(w_f)W_{\!\Delta}
(m_M\! -\! m_{\ell+1/2}^{(\ell)(n)})\nonumber\\
&&\mbox{}+ \sum_{\ell=1}S_{\ell}(w_i,w_f,w_{if})\!
\sum_{n} \tau_{\ell-1/2}^{(\ell)(n)}(w_i)
\tau_{\ell-1/2}^{(\ell)(n)}(w_f)W_{\!\Delta}
(m_M\! -\! m_{\ell-1/2}^{(\ell)(n)})\ ,
\end{eqnarray}
where the weight function now bounds excitation \emph{mass} because
$v'^0=1$. The functions $\tau_{\ell \pm 1/2}^{(\ell)(n)}(w)$ are
$\bar{B} \to D^{(\ell)(n)}_{\ell \pm 1/2} \,$ Isgur-Wise functions,
where $\ell\pm 1/2$ is the spin of the light degrees of freedom,
$(-1)^{\ell+1}$ is the parity of the state, and the label $n$ counts
``radial excitations.'' This name is inspired by the nonrelativistic
constituent quark model, but these states can be anything carrying the
other quantum numbers, including continuum contributions, for which
$n$ would be a continuous parameter and the sums integrals. In that
case, the functions $\tau_{\ell \pm 1/2}^{(\ell)(n)}(w)$ would not be
Isgur-Wise functions but other $B$ decay form factors. This
possibility will be downplayed here, with the assumption that such
contributions are small in the bounds derived here. Experimental input
on $B \to D \pi l \bar{\nu}$, for example, is needed to evaluate this
assumption.

The quark model also offers an interpretation of $\ell$ as the orbital
angular momentum between the light antiquark and the heavy quark. The
relation of this notation to that of Isgur and Wise~\cite{iwex} for
the lower values of $\ell$ is given by
\begin{equation}
\label{iwrel}
\begin{array}{r @{,} c @{,} l}
\tau_{1/2}^{(0)(n)}(w)=\xi^{(n)}(w) \quad & 
\quad \tau_{1/2}^{(1)(n)}(w)= 2 \, \tau_{1/2}^{(n)}(w) \quad & 
\quad \tau_{3/2}^{(1)(n)}(w)= \sqrt{3} \, \tau_{3/2}^{(n)}(w) \ .
\end{array}
\end{equation}
The function $S_\ell$ takes into account the polarization of an
intermediate state with integral spin $\ell$ and is defined as
\begin{equation}
S_\ell=v_{i \nu_1}\!\!\cdots v_{i \nu_\ell} 
v_{f \mu_1}\!\!\cdots v_{f \mu_\ell}\sum_\lambda
\varepsilon_{(\lambda)}^{* \nu_1 \cdots \nu_\ell}
\varepsilon_{(\lambda)}^{\mu_1 \cdots \mu_\ell}\ ,
\end{equation}
where $\varepsilon^{\mu_1 \cdots \mu_\ell}$ is the polarization tensor
of the intermediate state. The sum runs over the $2\ell+1$
polarizations. This quantity was reduced in Ref.~\cite{ley1} to the
relatively simple form
\begin{equation}
S_\ell(w_i,w_f,w_{if})= \sum_{k=0}^{\ell/2} C_{\ell,k}(w_i^2-1)^k (w_f^2-1)^k
(w_i w_f-w_{if})^{\ell-2k} \, ,
\end{equation}
with the coefficient
\begin{equation}
C_{\ell,k}= (-1)^k \frac{(\ell!)^2}{(2\ell)!} \frac{(2\ell-2k)!}
{k! (\ell-k)! (\ell-2k)!}\ .
\end{equation}
Using this formula it is easy to show that Eq.~(\ref{vector}) reduces
to
\begin{equation}
2(1+w)\xi(w)[1+\alpha_s F(1,w,w)]=2(1+w)\xi(w)
\end{equation}
in the limit $v_i=v'$, confirming that $F(1,w,w)=0$ to all orders. The
limit $v_f=v'$ gives $F(w,1,w)=0$. 

The axial sum rule (i.e., Eq.~(\ref{generic}) with $\Gamma_{\! i,f}={/
\hskip - 0.53em v}_{i,f} \gamma_5$) can be written out explicitly
in the same way~\cite{ley2}:
\begin{eqnarray}
\label{axial}
&& \xi(w_{if}) \left[1+ \alpha_s F(w_i,w_f,w_{if})\right]
(-1+w_i+w_f-w_{if}) \nonumber \\
&=& (w_i-1)(w_f-1)\!\sum_{\ell=1}\frac{\ell}{2\ell-1} 
S_{\ell-1}(w_i,w_f,w_{if})\!\sum_{n}\tau_{\ell-1/2}^{(\ell)(n)}(w_i)
\tau_{\ell-1/2}^{(\ell)(n)}(w_f)W_{\!\Delta}
(m_M\! -\! m_{\ell-1/2}^{(\ell)(n)})\nonumber\\
&&\mbox{}+ \sum_{\ell=0}S_{\ell+1}(w_i,w_f,w_{if})\!
\sum_{n} \tau_{\ell+1/2}^{(\ell)(n)}(w_i)
\tau_{\ell+1/2}^{(\ell)(n)}(w_f)W_{\!\Delta}
(m_M\! -\! m_{\ell+1/2}^{(\ell)(n)})\ .
\end{eqnarray}
As in the vector sum rule, the masses of the intermediate states are
denoted by $m_{\ell\pm 1/2}^{(\ell)(n)}$. The limits $v_i=v'$ and
$v_f=v'$ are trivial for the axial sum rule.

The doublets with spin of the light degrees of freedom $s_l=\ell+1/2$
and $s_l=\ell-1/2$ contain states with angular momentum $\ell,\,
\ell+1$ and $\ell-1,\, \ell$, respectively. But only one member of
each doublet contributes to the sum rules in Eqs.~(\ref{vector}) and
(\ref{axial}) because of the choice of currents. This explains the
appearance of only one polarization function for each doublet in the
vector and axial vector sum rules.

The zero-recoil normalization of the $\bar{B} \to D^{(*)}$
Isgur-Wise function allows one to write
\begin{equation}
\xi(w)=1-\rho^2(w-1)+\frac{\sigma^2}{2}(w-1)^2- \cdots\ .
\end{equation}
The axial and vector sum rules (i.e., Eqs.~(\ref{vector}) and
(\ref{axial})) give expressions for $\rho^2$, $\sigma^2$, and higher
derivatives of $\xi(w)$ when differentiated with respect to the
parameters $w_x$ at $w_x=1$. Different combinations of derivatives
yield different relations. In the $v'^0=1$ frame, the sum rules are
invariant under the interchange of $w_i$ and $w_f$, and it is
therefore sufficiently general to consider only derivatives with
respect to $w_{if}$ and $w=w_i=w_f$. Because of this simplification,
this paper only uses derivatives of the vector and axial sum rules of
the sort
\begin{equation}
\left. \frac{\partial^{p+q}}{\partial w_{if}^p \partial w^q}\,
\right|_{w_{if}=w=1} \ .
\end{equation}
Derivatives of the vector sum rule with $p+q=2$ give expressions for
$\sigma^2$, while the extra factors of $(w_x-1)$ in the axial rule
require $p+q=3$ for curvature relations.

As an illustration of the method, one can easily derive the
Bjorken~\cite{bjorken,iwex} and Uraltsev~\cite{uraltsev} sum rules
with order $\alpha_s$ corrections. For this it is only necessary to
consider $p+q=1$ in the vector rule and $p+q=2$ in the axial
rule. Taking the vector sum rule first, the equation given by the
$p=0,\, q=1$ derivative is trivial, but $p=1,\, q=0$ gives the Bjorken
sum rule with one-loop corrections~\cite{bjo}:
\begin{equation}
\label{bjsum}
\rho^2(\mu)=\frac{1}{4}+\frac{4\alpha_s}{9\pi}\left(\frac{5}{3}-
\mathrm{ln}\frac{4\Delta^2}{\mu^2}\right)
+2\sum_n^{\Delta}\left[\tau_{3/2}^{(n)}(1)\right]^2
+\sum_n^{\Delta}\left[\tau_{1/2}^{(n)}(1)\right]^2 \ .
\end{equation}
This equation has been written in the familiar notation of Isgur and
Wise using Eq.~(\ref{iwrel}). The upper limit $\Delta$ on the sums
stands for a factor of the weight function $W_{\!\Delta}(m_M\! -\!
m_{X_c})$, which serves to cut off the sums. Without it the results
are divergent, as can be seen by attempting to take the $\Delta \to
\infty$ limit in the order $\alpha_s$ corrections. Note that he
subtraction-point dependence is the same on both sides of the
equation, since Isgur-Wise functions are independent of $\mu$ at zero
recoil while their slopes depend on it
logarithmically~\cite{neubert}. This equation should be evaluated near
$\mu=\Delta$ to avoid large logarithms in the perturbative expansion.

The lower bound resulting from ignoring the sums in Eq.~(\ref{bjsum})
is similar to one derived in Ref.~\cite{competbd} but somewhat
weaker. As discussed in Ref.~\cite{bjo}, this is the result of using
different weight functions. That of Ref.~\cite{competbd} is
effectively given by the phase space of $b$ decay and falls off faster
with $\varepsilon$, thus reducing the contribution of the intermediate
states to the sum rule and strengthening the resultant lower bound. A
similar effect could be achieved here by using a smaller value for
$\Delta$, but as discussed above this makes the use of perturbative
QCD less reliable.

The $p=0,\, q=2$ derivative of the axial equation also gives the
Bjorken sum rule. The $p=2,\, q=0$ and $p=1,\, q=1$ derivatives give
the same result, which, when combined with the Bjorken rule, gives the
traditional form of the Uraltsev sum rule:
\begin{equation}
\sum_n^{\Delta}\left[\tau_{3/2}^{(n)}(1)\right]^2
-\sum_n^{\Delta}\left[\tau_{1/2}^{(n)}(1)\right]^2
=\frac{1}{4}-b_{if1}-\frac{1}{2}(b_{i1}+b_{f1})
=\frac{1}{4}\ ,
\end{equation}
where Eq.~(\ref{iwrel}) has again been used. This equation receives no
unsuppressed perturbative corrections. (There are in fact perturbative
corrections suppressed by
$\Lambda_{\mathrm{QCD}}^2/\Delta^2$~\cite{uraltsev}, but such
corrections are being neglected here.) In this particular derivation
of the Uraltsev rule, this is the result of the relation in
Eq.~(\ref{identb1}) between the first derivatives of $F$. But another
derivation from different derivatives of the axial and vector sum
rules gives $\alpha_s$ corrections proportional to $F(1,1,1)=0$. It
appears that the Uraltsev rule is always protected from perturbative
corrections by the general identities $F(1,w,w)=F(w,1,w)=0$. This
convergent sum rule indicates that $\tau_{1/2}^{(n)}(1)$ and
$\tau_{3/2}^{(n)}(1)$ become equal as $n \to \infty$.

Combined with Eq.~(\ref{bjsum}), the Uraltsev rule improves the
Bjorken bound significantly:
\begin{equation}
\label{uralt}
\rho^2(\mu)=\frac{3}{4}+b_{if1}(\mu)
+3\sum_n^{\Delta}\left[\tau_{1/2}^{(n)}(1)\right]^2
>\frac{3}{4}+\frac{4\alpha_s}{9\pi}\left(\frac{5}{3}-
\mathrm{ln}\frac{4\Delta^2}{\mu^2}\right)\ .
\end{equation}
Because the Uraltsev rule is not corrected, the corrections to this
improved bound are just those of the original Bjorken bound. In
particular, they are not substantially increased, as one might expect
from the drastic improvement to the bound.

Constraints on the curvature of the Isgur-Wise function are obtained
from higher derivatives of the same equations. The three second
derivatives of the vector equation and the four third derivatives of
the axial can be reduced to five linearly independent relations, as
demonstrated in Ref.~\cite{ley3}. Complete with the one-loop
corrections derived here, they are
\begin{eqnarray}
\rho^2&=& -\frac{4}{5}\sum_n^{\Delta}\tau_{3/2}^{(1)(n)}(1)
\tau_{3/2}^{(1)(n)'}(1)+\frac{3}{5}\sum_n^{\Delta}
\tau_{1/2}^{(1)(n)}(1) \tau_{1/2}^{(1)(n)'}(1)+\frac{2}{5}b_{if1}\ ,\\
\sigma^2&=& -\sum_n^{\Delta}\tau_{3/2}^{(1)(n)}(1) 
\tau_{3/2}^{(1)(n)'}(1)+b_{if1}\rho^2-b_{if2}
-\frac{1}{2}\left(b_{i2if}+b_{f2if}\right)\ ,\\
\sigma^2&=& 2\sum_n^{\Delta}\left[\tau_{5/2}^{(2)(n)}(1)\right]^2
+2b_{if1} \rho^2-b_{if2}\ ,\\
\sigma^2&=& \frac{5}{4}\rho^2+\frac{5}{4}
\sum_n^{\Delta}\left[\tau_{3/2}^{(2)(n)}(1)\right]^2+2b_{if1}\rho^2
-b_{if2}-\frac{5}{4}b_{if1}\ ,\\
\sigma^2&=& \frac{4}{5}\rho^2+\frac{3}{5}
\sum_n^{\Delta}\left[\tau_{1/2}^{(0)(n)'}(1)\right]^2+\frac{4}{5}b_{if1}\rho^2
-\frac{4}{5}b_{if1}-\frac{6}{5}\left(b_{i2if}+b_{f2if}\right) \nonumber \\
&&\mbox{}-\frac{8}{5}b_{if2}-\frac{3}{10}\left(b_{i2}+2b_{i2f}+b_{f2}\right)\ .
\end{eqnarray}
The last two equations give the bounds of Ref.~\cite{ley3}, complete
with order $\alpha_s$ corrections. Only a couple of orbital
excitations occur and in positive-definite form, allowing the
derivation of nontrivial lower bounds. Using the values of
Eqs.~(\ref{twoderiv}) gives
\begin{eqnarray}
\label{bdone}
\sigma^2(\mu)&>&\frac{5}{4}\rho^2(\mu)\left(1+\frac{32\alpha_s}
{27\pi}-\frac{32\alpha_s}{45\pi}\,\mathrm{ln}\frac{4\Delta^2}{\mu^2}\right)
-\frac{193\alpha_s}{675\pi}
+\frac{13\alpha_s}{45\pi}\,\mathrm{ln}\frac{4\Delta^2}{\mu^2}\ ,\\
\label{bdtwo}
\sigma^2(\mu)&>&\frac{3}{5}\left[\rho^2(\mu)\right]^2
+\frac{4}{5}\rho^2(\mu)\left(1+\frac{20\alpha_s}{27\pi}
-\frac{4\alpha_s}{9\pi}\,\mathrm{ln}\frac{4\Delta^2}{\mu^2}\right)
-\frac{148\alpha_s}{675\pi}
+\frac{4\alpha_s}{45\pi}\,\mathrm{ln}\frac{4\Delta^2}{\mu^2}\ ,
\end{eqnarray}
where the identity $\tau_{1/2}^{(0)(0)'}(1)=-\rho^2$ has been used. As
demonstrated below in the derivation of physical bounds, the
subtraction-point dependence is the same on both sides of these
inequalities.

\section{Physical Bounds}
\label{fo}

When combined with $\alpha_s$ and $\Lambda_{\mathrm{QCD}}/m_{c,b}$
corrections from matching HQET onto the full theory, the curvature
bounds derived above imply bounds on the zero-recoil derivatives of
the functions $\mathcal{F}_{D^{(*)}}(w)$. It is convenient to expand
these functions about the zero-recoil point according to
\begin{equation}
\label{expanf}
\mathcal{F}_{D^{(*)}}(w)=\mathcal{F}_{D^{(*)}}(1)
\left[1-\rho^2_{D^{(*)}}(w-1)+\frac{\sigma^2_{D^{(*)}}}{2}(w-1)^2-
\cdots \right] \ .
\end{equation}
A simple matching calculation, taken from Ref.~\cite{neubert}, yields
the relations between the Isgur-Wise derivatives and those of the
physical shape functions:
\begin{eqnarray}
\label{phys}
\rho_{D^{(*)}}^2&=&\rho^2(\mu)+\frac{4\alpha_s}{9\pi}\,\mathrm{ln}
\frac{m_c^2}{\mu^2}+\frac{\alpha_s}{\pi}
\left(\delta_{D^{(*)}}^{(\alpha_s)}-\frac{20}{27}\right)
+\frac{\bar{\Lambda}}{2m_c}\delta_{D^{(*)}}^{(1/m)}\ ,\nonumber \\
\sigma_{D^{(*)}}^2&=&\sigma^2(\mu)+2\rho^2(\mu)\left[\frac{4\alpha_s}{9\pi}\,
\mathrm{ln}\frac{m_c^2}{\mu^2}+\frac{\alpha_s}{\pi}
\left(\delta_{D^{(*)}}^{(\alpha_s)}-\frac{20}{27}\right)\right]
+\frac{4\alpha_s}{15\pi}\,\mathrm{ln}\frac{m_c^2}{\mu^2}\nonumber \\
&&\mbox{}+\frac{\alpha_s}{\pi}\left(\Delta_{D^{(*)}}^{(\alpha_s)}
-\frac{16}{25}\right)+\frac{\bar{\Lambda}}{2m_c}\Delta_{D^{(*)}}^{(1/m)}\ .
\end{eqnarray}
The perturbative corrections are model independent. The parameters
$\delta_{D^{(*)}}^{(\alpha_s)}$ and $\Delta_{D^{(*)}}^{(\alpha_s)}$
are given by
\begin{eqnarray}
\delta_{D^*}^{(\alpha_s)}&=&\frac{2(1-z)(11+2z+11z^2)+24(2-z+z^2)
z\,\mathrm{ln}z}{27(1-z)^3}=0.24\ ,\nonumber \\
\delta_D^{(\alpha_s)}&=&\frac{2(1-z)(23-34z+23z^2)+12(3-3z+2z^2)
z\,\mathrm{ln}z}{27(1-z)^3}=1.20\ ,\nonumber \\
\Delta_{D^*}^{(\alpha_s)}&=& -\frac{8(47+17z+252z^2+17z^3+47z^4)}
{675(1-z)^4}\nonumber\\
&&\mbox{}-\frac{4(5+125z-55z^2+95z^3-18z^4)z\,\mathrm{ln}z}
{135(1-z)^5}=-1.16\ ,\nonumber\\
\Delta_D^{(\alpha_s)}&=& \frac{4(47-258z+302z^2-258z^3+47z^4)}
{225(1-z)^4}\nonumber\\
&&\mbox{}-\frac{8(5-5z+5z^2-z^3)z^2\,\mathrm{ln}z}{15(1-z)^5}=0.63\ ,
\end{eqnarray}
where $z=m_c/m_b$, and the approximation $r_{(*)}\approx z$ has been
made in the order $\alpha_s$ corrections. These values agree with
those calculated in Ref.~\cite{grinlig}. The numerical values are for
$m_c=1.4$~GeV and $m_b=4.8$~GeV.

The nonperturbative corrections cannot currently be calculated
model-independently because they depend on the four subleading
Isgur-Wise functions that parameterize first-order corrections to the
heavy-quark limit, $\chi_{1-3}(w)$ and $\eta(w)$. But they can be
estimated using QCD sum rules~\cite{qcdsum} (and, in principle,
lattice QCD)\@. In the notation of Neubert~\cite{neubert}, the
nonperturbative corrections are
\begin{eqnarray}
\delta_{D^*}^{(1/m)}&=&-2\chi_1'(1)(1+z)-\frac{4}{3}\chi_2(1)(1-3z)
+4\chi_3'(1)(1-3z) \nonumber \\
&&\mbox{}-\frac{5}{6}(1+z)-\frac{1-2z+5z^2}{3(1-z)}\eta(1)
\approx -2.1\ ,\nonumber \\
\delta_D^{(1/m)}&=&-2(1+z)\left[\chi_1'(1)-2\chi_2(1)
+6\chi_3'(1)\right]+\frac{2(1-z)^2}{1+z}\eta'(1)\approx -1.3\ ,\nonumber\\
\Delta_{D^*}^{(1/m)}&=&\rho^2\left[-2\eta(1)\frac{1-2z+5z^2}{3(1-z)}
-\frac{5}{3}(1+z)\right]+2(1+z)\chi_1''(1)-\frac{8(1-6z+z^2)\chi_2(1)}
{9(1-z)^2}\nonumber \\
&&\mbox{}+\frac{8}{3}(1-3z)\chi_2'(1)-4(1-3z)\chi_3''(1)
-\frac{\eta(1)(5-28z+18z^2-52z^3+25z^4)}{9(1-z)^3}\nonumber \\
&&\mbox{}+\frac{2\eta'(1)(1-2z+5z^2)}{3(1-z)}
-\frac{(1+z)(25-42z+25z^2)}{18(1-z)^2}\approx -2.6\rho^2-1.7\ ,\nonumber\\
\Delta_D^{(1/m)}&=&4\rho^2\eta'(1)\frac{(1-z)^2}{1+z}+2(1+z)\left[\chi_1''(1)
-4\chi_2'(1)+6\chi_3''(1)\right]-2\eta''(1)\frac{(1-z)^2}{1+z}\nonumber \\
&\approx& -0.3\ ,
\end{eqnarray}
where the primes denote $d/dw$. In these corrections $\rho^2$ can be
taken to be $\rho_{D^{(*)}}^2$, since the results here do not include
corrections of order $\alpha_s \Lambda_{\mathrm{QCD}}/m_{c,b}$. The
parts of these expressions for $\Delta_{D^*}^{(1/m)}$ and
$\Delta_D^{(1/m)}$ proportional to $\rho^2$ disagree with those of
Ref.~\cite{grinlig}.\footnote{The authors of Ref.~\cite{grinlig} have
confirmed these findings. They report that the numerical result in
their Eq.~(19) for the difference $(\sigma_D^2-\sigma_{D^*}^2)/2$
changes from $0.17+0.20\rho^2$ to $0.17+0.29\rho^2$.} The numerical
estimates are based on the approximate values $\eta(1)=0.6$,
$\eta'(1)=0$, $\chi_1'(1)=0.3$, $\chi_2(1)=-0.04$, $\chi_2'(1)=0.03$,
and $\chi_3'(1)=0.02$~\cite{qcdsum}. The values
$\eta''(1)=\chi_1''(1)=\chi_3''(1)=0$ and $z=0.29$ were also
used. Since these values are model dependent with large uncertainties,
the numerical estimates of the nonperturbative corrections should be
interpreted with caution. Reliable lattice results would be a welcome
check on such large QCD sum rule estimates of these corrections.

Combining the bounds of Eqs.~(\ref{bdone}) and (\ref{bdtwo}) with
Eqs.~(\ref{phys}) gives the physical bounds
\begin{eqnarray}
\label{phys1}
\sigma_{D^{(*)}}^2 &>& \frac{5}{4}\rho_{D^{(*)}}^2\left(1+
\frac{8\alpha_s}{5\pi}\delta_{D^{(*)}}^{(\alpha_s)}+
\frac{32\alpha_s}{45\pi}\,\mathrm{ln}\frac{m_c^2}{4\Delta^2}\right)
-\frac{13\alpha_s}{45\pi}\,\mathrm{ln}\frac{m_c^2}{4\Delta^2}
-\frac{5\alpha_s}{4\pi}\delta_{D^{(*)}}^{(\alpha_s)}
+\frac{\alpha_s}{\pi}\Delta_{D^{(*)}}^{(\alpha_s)}\nonumber\\
&&\mbox{}-\frac{5}{4}\frac{\bar{\Lambda}}{2m_c}\delta_{D^{(*)}}^{(1/m)}
+\frac{\bar{\Lambda}}{2m_c}\Delta_{D^{(*)}}^{(1/m)}\ ,\\
\label{phys2}
\sigma_{D^{(*)}}^2 &>&\frac{3}{5}\left(\rho_{D^{(*)}}^2\right)^2
+\frac{4}{5}\rho_{D^{(*)}}^2\left(1+\frac{4\alpha_s}{9\pi}\,
\mathrm{ln}\frac{m_c^2}{4\Delta^2}+\frac{\alpha_s}{\pi}
\delta_{D^{(*)}}^{(\alpha_s)}-\frac{3}{2}
\frac{\bar{\Lambda}}{2m_c}\delta_{D^{(*)}}^{(1/m)}\right)
-\frac{4\alpha_s}{45\pi}\,\mathrm{ln}\frac{m_c^2}{4\Delta^2}\nonumber\\
&&\mbox{}-\frac{4\alpha_s}{15\pi}-\frac{4\alpha_s}{5\pi}
\delta_{D^{(*)}}^{(\alpha_s)}+\frac{\alpha_s}{\pi}
\Delta_{D^{(*)}}^{(\alpha_s)}-\frac{4}{5}
\frac{\bar{\Lambda}}{2m_c}\delta_{D^{(*)}}^{(1/m)}
+\frac{\bar{\Lambda}}{2m_c}\Delta_{D^{(*)}}^{(1/m)}\ .
\end{eqnarray}
Numerically, these inequalities are
\begin{eqnarray}
\sigma_{D^*}^2&>&\frac{5}{4}\rho_{D^*}^2[1-0.11(0.16)_p-0.3_{np}]
-0.081(0.059)_p+0.1_{np}\ ,\nonumber\\
\sigma_{D^*}^2&>&\frac{3}{5}\left(\rho_{D^*}^2\right)^2
+\frac{4}{5}\rho_{D^*}^2[1-0.066(0.101)_p+0.08_{np}]
-0.14(0.13)_p-0.003_{np}\ ,\nonumber\\
\sigma_D^2&>&\frac{5}{4}\rho_D^2[1+0.041(-0.014)_p+0.0_{np}]
-0.025(0.0028)_p+0.2_{np}\ ,\nonumber\\
\sigma_D^2&>&\frac{3}{5}\left(\rho_D^2\right)^2
+\frac{4}{5}\rho_D^2[1+0.025(-0.0089)_p+0.3_{np}]-0.039(0.032)_p+0.1_{np} \ ,
\end{eqnarray}
where the values $\alpha_s=0.3$ (in the $\overline{\mathrm{MS}}$
scheme around 2 GeV) and $\bar{\Lambda}=0.4$ GeV have been used. The
perturbative corrections, with subscript $p$, are for two values of
$\Delta$. The results for $\Delta=2$ GeV are first, and those for
$\Delta=3$ GeV are in parentheses. The nonperturbative corrections are
labeled by a subscript $np$.

Equations~(\ref{phys1}) and (\ref{phys2}) imply absolute bounds when
combined with the corrected form of the Uraltsev bound, 
\begin{equation}
\label{absoura}
\rho_{D^{(*)}}^2>\frac{3}{4}+\frac{4\alpha_s}{9\pi}\,
\mathrm{ln}\frac{m_c^2}{4\Delta^2}+\frac{\alpha_s}{\pi}
\delta_{D^{(*)}}^{(\alpha_s)}+
\frac{\bar{\Lambda}}{2m_c}\delta_{D^{(*)}}^{(1/m)}\ ,
\end{equation}
which comes from Eq.~(\ref{uralt}) and the first of Eqs.~(\ref{phys}),
and the tree-level Voloshin bound~\cite{volosh}, $\rho^2 \lesssim
3/4$. A lower bound is required for terms proportional to
$\rho_{D^{(*)}}^2$ with positive coefficients, and an upper bound is
required for those with negative coefficients. The latter are
corrections, so the upper bound is only needed at tree level. Note
that an upper bound is required to estimate the greatest impact the
corrections could have on the bound. 

Inserting these inequalities into Eq.~(\ref{phys1}) gives
\begin{equation}
\label{abso}
\sigma_{D^{(*)}}^2 > \frac{15}{16}+\frac{14\alpha_s}{15\pi}\,
\mathrm{ln}\frac{m_c^2}{4\Delta^2}+\frac{3\alpha_s}{2\pi}
\delta_{D^{(*)}}^{(\alpha_s)}+\frac{\alpha_s}{\pi}
\Delta_{D^{(*)}}^{(\alpha_s)}+\frac{\bar{\Lambda}}{2m_c}
\Delta_{D^{(*)}}^{(1/m)}\ ,
\end{equation}
where $\rho_{D^{(*)}}^2$ is replaced by $3/4$ in
$\Delta_{D^{(*)}}^{(1/m)}$. The absolute bound produced in the same
way from Eq.~(\ref{phys2}) happens to be identical at leading order
and weaker only by the addition of $-4\alpha_s/15\pi$ after
perturbative and $\Lambda_{\mathrm{QCD}}/m_{c,b}$ corrections are
included. Using the numerical estimates above, the bounds in
Eq.~(\ref{abso}) are
\begin{eqnarray} 
\sigma_{D^*}^2&>&0.94-0.26(0.34)_p-0.5_{np}\ ,\nonumber \\
\sigma_D^2&>&0.94+0.045(-0.027)_p-0.04_{np} \ ,
\end{eqnarray}
where the corrections are labeled as described above.

When considering the bounds in Eqs.~(\ref{absoura}) and (\ref{abso})
and their dependence on $\Delta$, one must bear in mind that the
logarithms in the perturbative corrections are only small if $\Delta$,
$m_b$, and $m_c$ are roughly of the same order. That is the accuracy
of the results obtained here. For instance, Eq.~(\ref{phys}) is valid
for $\mu$ on the order of $m_{c,b}$, while Eqs.~(\ref{bdone}) and
(\ref{bdtwo}) are valid for $\mu$ near $\Delta$. Therefore, taking the
$\Delta \to \infty$ limit does not make sense in the absolute
bounds. To understand the behavior of the bounds in this limit, one
would need to sum the logarithms of $m_c^2/\Delta^2$. Since these
logarithms are not large for the values of $\Delta$ used here, this
extra step has been omitted.

\begin{figure}[b]
\begin{center}
\includegraphics[width=0.6\textwidth]{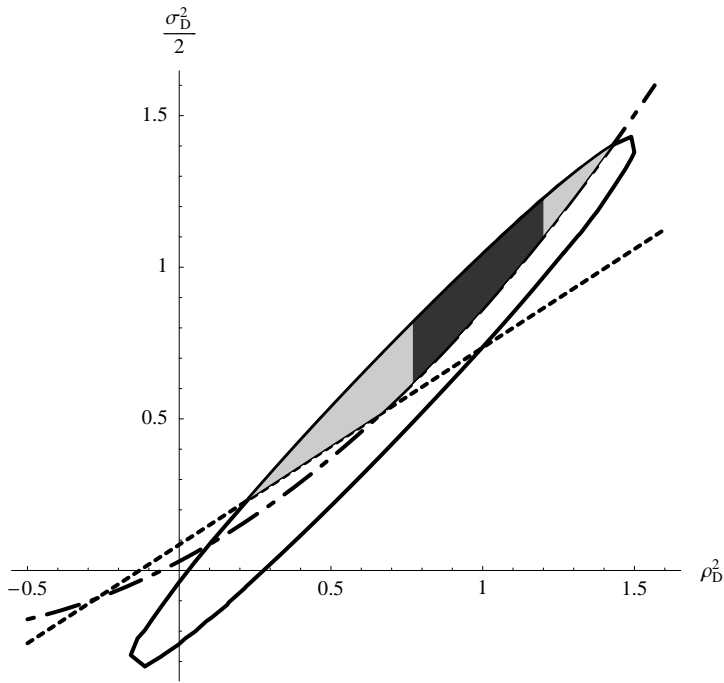}
\caption{\small Dispersive constraints on $\mathcal{F}_D$ derivatives
  combined with the corrected sum rule bounds derived here at
  $\Delta=2$ GeV. The interior of the ellipse is the region allowed by
  the dispersion relations. Including the curvature bounds, given by
  the area above the dashed curves, further restricts the allowed
  region to the shaded area. The darker region is obtained by also
  including the Bjorken and Voloshin bounds. Both perturbative and
  nonperturbative corrections are included.}
\label{compfig}
\end{center}
\end{figure}
The sum rule bounds derived here should be compared with the
dispersive constraints usually used to guide the extrapolation of
measured form factors to zero recoil. These constraints are derived by
computing the vacuum expectation value of a time-ordered product of $b
\to c$ currents in the perturbative regime and then using analyticity
to learn about the semileptonic regime. The result is equated with a
spectral function sum of resonances (i.e., a sum of positive
quantities). Much as in the derivation of the sum rules here, focusing
on specific resonances yields form factor constraints. A typical
example is shown in Fig.~\ref{compfig}. The slope and curvature must
lie within the ellipse, a constraint that is well approximated by a
linear relation between the slope and curvature. Data are fit as a
function of $w$ for $|V_{cb}| \mathcal{F}_{D^{(*)}}(1)$ and
$\rho_{D^{(*)}}^2$, and the second derivative at zero recoil is
related to the slope according to this relation. For the process
$\bar{B}^0 \to D^+ \ell^-\bar{\nu}$, Belle used the relation
$\sigma_D^2/2=1.05\rho_D^2-0.15$~\cite{disper2} to find
$\sigma_D^2=2.06 \pm 0.46 \pm 0.29$, where the first uncertainty is
statistical and the second systematic~\cite{belle1}. This value is
consistent with the bound above.

Rather than $\mathcal{F}_{D^*}(w)$, one typically fits the shape of
the axial vector form factor $h_{A_1}(w)$, which is defined, for
example, in Ref.~\cite{grinlig}. Like $\mathcal{F}_{D^*}(w)$, it is
equal to the Isgur-Wise function $\xi(w)$ in the heavy-quark
limit. Its curvature is defined as in Eq.~(\ref{expanf}) and satisfies
the bound in Eq.~(\ref{abso}), with perturbative and nonperturbative
corrections given by
\begin{eqnarray}
\delta_{A_1}^{(\alpha_s)}&=&\frac{2(1-z)(17-4z+17z^2)
+6(9-3z+4z^2)z\,\mathrm{ln}z}{27(1-z)^3}=0.65\ ,\nonumber \\
\delta_{A_1}^{(1/m)}&=&-2(1+z)\chi_1'(1)+4z\chi_2(1)
+4\chi_3'(1)(1-3z)+z\eta(1)-\frac{1+z}{2}\approx -1.3\ ,\nonumber \\
\Delta_{A_1}^{(\alpha_s)}&=&\frac{4(1-z)(27-203z-68z^2-203z^3+27z^4)
-120(10+5z^2-z^3)z^2\,\mathrm{ln}z}{225(1-z)^5}=0.24\ ,\nonumber \\
\Delta_{A_1}^{(1/m)}&=&\rho^2[2z\eta(1)-1-z]+2\chi_1''(1)(1+z)
-8z\chi_2'(1)-4\chi_3''(1)(1-3z)\nonumber \\
&&\mbox{}+z\eta(1)-2z\eta'(1)-\frac{1+z}{2}\approx -0.9\rho^2-0.5\ ,
\end{eqnarray}
where the numerical estimates are based on the values used
above. These values produce the absolute bound
\begin{equation}
\sigma_{A_1}^2 > .94-0.071(0.14)_p-0.2_{np}
\end{equation}
on the curvature of $h_{A_1}(w)$. This should be compared with the
value found by Belle by the procedure described above for the process
$\bar{B}^0 \to D^{* +} e^- \bar{\nu}$. Using the relation
$\sigma_{A_1}^2/2=1.08\rho_{A_1}^2-0.23$~\cite{disper2}, Belle found
$\sigma_{A_1}^2=2.44 \pm 0.37 \pm 0.41$, where the first uncertainty
is statistical and the second systematic~\cite{belle2}. This value is
also consistent with the bound produced here. A plot comparing
dispersive constraints and sum rule bounds for this form factor would
be similar to Fig.~\ref{compfig}, but with the sum rule bounds
comparatively somewhat weaker.

\section{Conclusions}

This paper has presented order $\alpha_s$ corrections to two new sum
rules derived in Refs.~\cite{ley1,ley2,ley3} in the context of
HQET. Section~\ref{tw} repeated the tree-level derivation of a generic
sum rule depending on three velocity transfer variables and included
one-loop corrections. Section~\ref{th} studied the axial vector and
vector sum rules that result from choosing specific currents in the
generic equation. These led to $\alpha_s$-corrected versions of the
sum rules of Le Yaouanc \emph{et al.}\ for the curvature of the
Isgur-Wise function. There were no corrections suppressed by the heavy
quark masses because the infinite-mass limit was
used. Section~\ref{fo} translated these HQET bounds into bounds on
physical form factors by including perturbative and finite-mass
corrections associated with matching HQET onto the full
theory. Numerical estimates were given and compared with experimental
values and dispersive constraints. The bounds produced here are less
powerful than dispersive constraints but may provide an important
check on those constraints.

\acknowledgments 
I would like to thank Mark Wise for many discussions and a critical
reading of the draft, and Zoltan Ligeti for several helpful comments
on the draft. This work was supported in part by a National Science
Foundation Graduate Research Fellowship and the Department of Energy
under Grant No. DE-FG03-92ER40701.

\end{document}